\documentclass[a4paper,12pt, epsfig]{article}
\def\letter{0}\def\pr{0}
\pdfoutput=1 
\usepackage{epsfig}
\usepackage{epstopdf}
\usepackage{graphicx}
\usepackage{ifthen}

\usepackage{amsmath}
\usepackage[psamsfonts]{amssymb}
\usepackage{euscript}

\usepackage{latexsym}
\usepackage[arrow,matrix,curve]{xy}

\newskip\humongous \humongous=0pt plus 1000pt minus 1000pt

\newif\ifdtup

\def\,{\hspace{-.1cm}}
\def\hsp{,\hspace{.7cm}}

\def\tf {\tilde{f}}
\def\ff{{\mathcal{F}}}

\def\fc#1#2 {\frac{n}{q}#1\frac{n}{q}#2}

\def\hf{H\p_{f_0}}
\def\hpt{H\p_{\rm{free}}}

\newcommand{\vac}{\ensuremath{|0\rangle}}
\newcommand{\stt}{\ensuremath{|\psi\rangle}}

\renewcommand{\cos}{\textrm{cos}}
\renewcommand{\sin}{\textrm{sin}}

\renewcommand{\sinh}{\textrm{sinh}}
\renewcommand{\cosh}{\textrm{cosh}}
\renewcommand{\tanh}{\textrm{tanh}}
\newcommand{\sech}{\textrm{sech}}
\newcommand{\csch}{\textrm{csch}}

\def\exp#1{\hbox{\rm exp}\left(#1\right)}

\renewcommand{\theequation}{\arabic{section}.\arabic{equation}}
\renewcommand{\(}{\begin{equation}}
\renewcommand{\)}{end{equation} \vspace{-.05in}\linebreak}

\newcounter{saveeqn}
\newcounter{savealpheqn}

\newcommand{\alpheqn}{\setcounter{saveeqn}{\value{equation}}%
  \stepcounter{saveeqn}\setcounter{equation}{0}%
  \renewcommand{\theequation}{\mbox{\arabic{section}.\arabic{saveeqn}
\alph{equation}}}
  \renewcommand{\)}{\end{equation}}}
\def\part#1{\frac{\partial}{\partial{#1}}}%
\def\group#1{\refstepcounter{equation}\setcounter{saveeqn}
 {\value{equation}}%
  \label{#1}\setcounter{equation}{0}%
\renewcommand{\theequation}{\mbox{\arabic{section}.\arabic{saveeqn}
\alph{equation}}}
  \renewcommand{\)}{\end{equation}}}
\newcommand{\reseteqn}{\setcounter{equation}{\value{saveeqn}}%
  \renewcommand{\theequation}{\arabic{section}.\arabic{equation}}%
  \renewcommand{\)}{\end{equation}}}

\newcommand{\aalpheqn}{\setcounter{saveeqn}{\value{equation}}%
  \stepcounter{saveeqn}\setcounter{equation}{0}%
  \renewcommand{\theequation}{\mbox{
        \Alph{subsection}.\arabic{saveeqn}\alph{equation}}}
   \renewcommand{\)}{\end{equation}}}
\newcommand{\areseteqn}{\setcounter{equation}{\value{saveeqn}}%
  \renewcommand{\theequation}{\Alph{subsection}.\arabic{equation}}%
  \renewcommand{\)}{\end{equation}}}

\renewcommand{\thefootnote}{\alph{footnote}}
\renewcommand{\(}{\begin{equation}}
\renewcommand{\)}{\end{equation}}
\newcommand{\ba}{\begin{eqnarray}}
\newcommand{\ea}{\end{eqnarray}}

\renewcommand{\b}{\beta}

\newcommand{\vt}{V^{(3)}(\sqrt{\lambda}f(x))}
\newcommand{\cbp}{\mathop{\vtop{\ialign{##\crcr
   $\hfil\displaystyle{}\hfil$\crcr\noalign{\kern-13pt\nointerlineskip}
   \BIG{)}\hskip0pt\crcr\noalign{\kern3pt}}}}}
\newcommand{\pa}{\mathop{\vtop{\ialign{##\crcr

$\hfil\displaystyle{\oplus}\hfil$\crcr\noalign{\kern+1pt\nointerlineskip
}
   \hspace{.08in}$^{\alpha=0}$\hskip6pt\crcr\noalign{\kern3pt}}}}}
\renewcommand{\hsp}{,\hspace{.3in}}
\newcommand{\p}{^\prime}
\newcommand{\pp}{^{\prime\prime}}

\def\D{\ensuremath{{\cal D}}}

\catcode`\@=11
\def\vereq#1#2{\lower3pt\vbox{\baselineskip1.5pt \lineskip1.5pt
\ialign{$\m@th#1\hfill##\hfil$\crcr#2\crcr\sim\crcr}}}
\catcode`\@=12

\renewcommand{\(}{\begin{equation}}
\renewcommand{\)}{\end{equation}}

\def\pin#1{\int \frac{d#1}{2\pi}}
\def\ppin#1{\int\hspace{-17pt}\sum \frac{d#1}{2\pi}}
\def\ppink#1{\int\hspace{-17pt}\sum\frac{d^{#1}k}{(2\pi)^{#1}}}
\def\dint{\int\hspace{-12pt}\sum }
\def\pink#1{\int \frac{d^{#1}k}{(2\pi)^{#1}}}

\def\Bd#1{B^\dag_{k_{#1}}}

\def\tp#1#2#3{\hbox{\rm tan}^#1(\thet#2#3)}

\def\cc{\mathcal{C}}
\def\df{\mathcal{D}_{f}}

\def\dF{\mathcal{D}_F}

\def\I{\mathcal{I}}

\def\os{\omega_S}

\def\as{|\alpha;\sigma\rangle}
\def\asb{\langle\alpha;\sigma|}

\newcommand{\beas}{\begin{eqnarray*}}
\newcommand{\eeas}{\end{eqnarray*}}

\newcommand{\bquo}{\begin{quote}}
\newcommand{\enqu}{\end{quote}}

\newcommand{\g}{{\mathfrak g}}

\def\ch{{\mathcal{H}}}

\def\tp{{\tilde{\phi}}}
\def\ok#1{\omega_{k_{#1}}}

\def\V#1{V^{(#1)}(\sqrt{\lambda}f(x))}

\def\ck{\csch\left(\frac{\pi k}{2\beta}\right)}

\newcommand{\beq}{\begin{equation}}
\newcommand{\eeq}{\end{equation}}
\newcommand{\bea}{\begin{eqnarray}}
\newcommand{\eea}{\end{eqnarray}}

\newskip\humongous \humongous=0pt plus 1000pt minus 1000pt

\newif\ifdtup

\catcode`\@=11

\ifthenelse{\equal{\letter}{0}}{ 

\usepackage[hypertexnames=false]{hyperref}
\hypersetup{
    colorlinks=true,
    linkcolor=blue,
    filecolor=magenta,      
    urlcolor=cyan,
    pdftitle={Overleaf Example},
    pdfpagemode=FullScreen,
    }

\urlstyle{same}

\@addtoreset{equation}{section}
\def\theequation{\arabic{section}.\arabic{equation}}

\def\@normalsize{\@setsize\normalsize{15pt}\xiipt\@xiipt
\abovedisplayskip 14pt plus3pt minus3pt%
\belowdisplayskip \abovedisplayskip
\abovedisplayshortskip \z@ plus3pt%
\belowdisplayshortskip 7pt plus3.5pt minus0pt}

\def\small{\@setsize\small{13.6pt}\xipt\@xipt
\abovedisplayskip 13pt plus3pt minus3pt%
\belowdisplayskip \abovedisplayskip
\abovedisplayshortskip \z@ plus3pt%
\belowdisplayshortskip 7pt plus3.5pt minus0pt
\def\@listi{\parsep 4.5pt plus 2pt minus 1pt
      \itemsep \parsep
      \topsep 9pt plus 3pt minus 3pt}}

\relax

\def\section{\@startsection{section}{1}{\z@}{3.5ex plus 1ex minus  .2ex}{2.3ex plus .2ex}{\large\bf}}

\def\thesection{\arabic{section}}
\def\thesubsection{\arabic{section}.\arabic{subsection}}

\def\appendix{\setcounter{section}{0}
 \def\thesection{Appendix \Alph{section}}
 \def\thesubsection{\Alph{section}.\arabic{subsection}}
 \def\theequation{\Alph{section}.\arabic{equation}}}
\renewcommand{\theequation}{\arabic{section}.\arabic{equation}}

}{
\renewcommand{\theequation}{\arabic{equation}}

} 

\begin{document}
\def\thefootnote{\fnsymbol{footnote}}
\def\thetitle{Spontaneous Emission from Excited Quantum Kinks}
\def\autone{Jarah Evslin}
\def\auttwo{Alberto Garc\'ia Mart\'in-Caro}
\def\affa{Institute of Modern Physics, NanChangLu 509, Lanzhou 730000, China}
\def\affb{University of the Chinese Academy of Sciences, YuQuanLu 19A, Beijing 100049, China}
\def\affc{Departamento de F\'isica de Part\'iculas, Universidad de
Santiago de Compostela and \\
Instituto Galego de F\'isica de Altas Enerxias
, E-15782
Santiago de Compostela, Spain}
\def\affd{Physics Dept., Brookhaven National Laboratory, Bldg. 510A, Upton, NY 11973, USA}

\title{Titolo}

\ifthenelse{\equal{\pr}{1}}{
\title{\thetitle}
\author{\autone}
\affiliation {\affa}
\affiliation {\affb}
\affiliation {\affc}
\affiliation {\affd}

}{

\begin{center}
{\large {\bf \thetitle}}

\bigskip

\bigskip


{\large \noindent  \autone{${}^{1,2}$} \footnote{jarah@impcas.ac.cn} 
and \auttwo{${}^{3,4}$} \footnote{alberto.martincaro@usc.es}}


\vskip.7cm

1) \affa\\
2) \affb\\
3) \affc\\
4) \affd\\

\end{center}

}

\begin{abstract}
\noindent
Many kink solutions enjoy internal excitations, called shape modes.  In some 1+1d scalar models, such as the $\phi^4$ double-well model, when a kink's shape mode is excited twice it may decay to a ground state kink plus a meson.  We analytically calculate the decay rates of both the twice-excited shape mode and also a coherent state corresponding to the classically excited shape mode.  In the case of the $\phi^4$ model, we find that the latter agrees with the classical result of Manton and Merabet.

\end{abstract}

%
\setcounter{footnote}{0}
\renewcommand{\thefootnote}{\arabic{footnote}}

\ifthenelse{\equal{\pr}{1}}
{
\maketitle
}{}

\section{Introduction}

The spectrum of a (1+1)-dimensional scalar quantum field theory whose potential has degenerate minima has been understood for half a century \cite{dhn2}.  It contains a plane wave representing a monochromatic elementary meson and a quantum kink with arbitrary momentum.  Integrable models contain arbitrary superpositions of these, while some superpositions are also in the spectrum of nonintegrable models such as excited normal modes in the one-kink sector.  In some nonintegrable models, the kink has localized excitations called shape modes which can be excited up to some maximum number of times.  For example, the $\phi^4$ double-well model contains a single shape mode which may be excited once.

While the spectrum of the Hamiltonian is fairly well understood, there are also a number of long-lived but ultimately unstable objects in these theories.  In all sectors, including the zero-kink sector, there can be long-lived oscillons \cite{osc,osc3d}.  These are large amplitude oscillations which, like the kinks themselves, are beyond the reach of perturbation theory.  In the multikink sectors and the kink-antikink sector there can be unstable collective states of the combined system \cite{doreyf6,col22}.  In the single kink sector, a shape mode may be excited multiple times, so that its energy is sufficient for it to escape into the meson continuum.

So far, all of these unstable but long-lived configurations have been studied almost exclusively in the classical theory.  One exception is Ref.~\cite{quantosc} where a semiclassical quantization was performed about a fixed oscillon solution background.  The resulting quantum corrections dramatically reduced the oscillon lifetime, as a result of perturbative decay channels which only exist in the quantum theory.

In the present note, we would like to understand the unstable multiply-excited shape mode in the full quantum field theory.  We will begin with the twice-excited shape mode.  Using the linearized kink perturbation theory of Refs.~\cite{mekink,me2loop} at tree level, we perform the first ever computation of its lifetime.  

As a consistency check, we will also consider a different regime.  We will excite the shape modes in a coherent state, which is not an eigenstate of the shape mode number operator and in general contains a number of shape modes which diverges at weak coupling.  This case is equivalent to the excited shape mode in classical field theory, whose decay was studied in Refs.~\cite{mm,mmnum}.  We will calculate the power radiated by a coherent state of shape modes, using the same process as is responsible for the decay in the doubly-excited case.  Our analytical result will agree precisely with the analytic result obtained years ago in classical field theory in Ref.~\cite{mm}, which in turn agrees, for small oscillation amplitudes, with the numerical results of Ref. \cite{mmnum}.

We begin in Sec.~\ref{revsez} with a review of linearized kink perturbation theory.  In Sec.~\ref{due} we compute the decay rate of a twice-excited shape mode.  Then in Sec.~\ref{tanti} we extend this computation to the coherent state, calculating the radiated power from the decays of pairs of shape modes as a function of the classical shape mode amplitude.  

\section{Review} \label{revsez}

While we believe that linearized kink perturbation theory can be generalized to more dimensions and to include fermions, so far all studies have concerned (1+1)d quantum field theories of a scalar field $\phi(x)$, whose elementary quanta will be referred to as mesons, and its conjugate momentum $\pi(x)$.  In particular, we consider the class of theories defined by a Hamiltonian $H$
\bea
H&=&\int dx :\mathcal{H}(x):_a \hsp \mathcal{H}(x)=\frac{\pi^2(x)}{2}+\frac{\left(\partial_x \phi(x)\right)^2}{2}+\frac{V(\sqrt{\lambda}\phi(x))}{\lambda}\nonumber\\ 
m^2&=&V^{(2)}(\sqrt{\lambda}\phi(\pm\infty))\hsp
V^{(n)}(\sqrt{\lambda}\phi(x))=\frac{\partial^n V(\sqrt{\lambda}\phi(x))}{(\partial \sqrt{\lambda}\phi(x))^n}.
\eea
Here $::_a$ is the normal ordering of plane wave creation and annihilation operators at a mass scale of $m$.  We will be interested in a semiclassical expansion in the parameter $\lambda$, which may or may not have nontrivial mass dimensions, so long as the potential $V/\lambda$ has the dimensions of [mass]${}^2$.  Such theories are particularly simple because the normal ordering removes all ultraviolet divergences.

We are interested in potentials $V$ with degenerate minima, so that the classical equations of motion will have stationary kink solutions $f$ which interpolate between the minima
\beq
\phi(x,t)=f(x)\hsp
-\sqrt{\lambda}f^{''}(x)+\V{1}=0
\label{fd}
\eeq
and have classical mass $Q_0$.  We choose one stationary kink solution $f(x)$, or more generally one point in the moduli space of solutions.  With this choice, the translation invariance of the theory is no longer manifest.  However, the translation invariance itself is not broken and so we expect it to appear in calculated observable quantities after summing to all orders in perturbation theory.

In the quantum theory, the kinks are represented by states.  These states do not appear in the usual Fock space of meson plane waves in the vacuum sector.  Instead they appear in the one-kink sector, which includes a kink plus an arbitrary number of internal and continuum normal mode excitations.  As this sector does not appear in the usual plane wave Fock space obtained by quantizing $H$, it lies beyond the reach of ordinary perturbation theory.

In the classical field theory this problem reflects the fact that the solution $\phi=f$ is far from the value $\phi=0$ about which one would expand in a perturbative approach.  This problem can be removed by replacing the Hamiltonian $H[\phi,\pi]$ with a kink Hamiltonian $H\p[\phi,\pi]=H[\phi+f,\phi]$.   The kink Hamiltonian yields the same mass spectra and evolution, but has the advantage that it allows classical field configurations close to the kink to be treated perturbatively.

One would like to apply the same trick to the quantum field theory.  However the Hamiltonian in quantum field theory is necessarily regularized, and there is no guarantee that a given regularization scheme will be compatible with such a transformation.  Indeed, naively replacing $\phi$ by $\phi+f$ in the Hamiltonian and then applying a chosen matching condition to the regulators of $H$ and $H\p$ leads to different mass spectra for the two Hamiltonians \cite{rebhan}.  Since the theory is defined by $H$, its mass spectrum is the correct one, and so masses calculated using $H\p$ are sometimes incorrect.

We thus need a more careful definition of $H\p$ in the quantum field theory.  Following Ref.~\cite{mekink}, we define the kink Hamiltonian $H\p$ using the unitary displacement operator $\df$
\beq
\df={{\rm Exp}}\left[-i\int dx f(x)\pi(x)\right]\hsp
H\p=\df^\dag H\df. \label{df}
\eeq
The displacement operator satisfies
\beq
:F[\phi(x),\pi(x)]:_a\df=\df:F[\phi(x)+f(x),\pi(x)]:_a
\eeq
and so corresponds to the classical notion of shifting the field $\phi(x)$ by $f(x)$.  However, by construction $H\p$ is unitarily equivalent to $H$, and so it will have the same spectrum, and therefore the correct spectrum.  Furthermore, as
\beq
\df e^{-iH\p t}\df^\dag=e^{-iHt}
\eeq
the kink Hamiltonian $H\p$ can be used to evolve the state, so long as one remembers to act with $\df^\dag$ before and $\df$ after.

This situation can be understood as follows.  The Hilbert space is defined in a given frame, where the operator $H$ generates time translations and its eigenvalue is the energy.  However $\df^\dag$ transforms it into a different frame, called the kink frame, where it is $H\p$ that defines time translations and whose eigenvalue is the energy.  Critically, these energies agree.  The advantage of linearized kink perturbation theory is that all calculations in the one-kink sector are performed in the kink frame, where they are perturbative.  Then, if desired, one can always transform states back to the defining sector by acting with $\df$.

Now that we have provided the basic definitions of this formalism, it remains to plug in and calculate \cite{zitto}.  For example, from Eq.~(\ref{df}) one may calculate the kink Hamiltonian $H\p$.  It will be convenient to decompose $H\p$ into summands $H\p_j$ which, when plane-wave normal-ordered, each contain $j$ factors of $\phi$ and $\pi$.   Then one finds
\bea
H\p_0&=&Q_0\hsp 
H\p_1=0\hsp
H\p_{n>2}=\lambda^{\frac{n}{2}-1}\int dx\frac{V^{(n)}(\sqrt{\lambda}f(x))}{n!}:\phi^n(x):_a\nonumber\\
H\p_2&=&\frac{1}{2}\int dx\left[:\pi^2(x):_a+:\left(\partial_x\phi(x)\right)^2:_a+\V{2}:\phi^2(x):_a\right].\label{h2}
\eea

Just as the free part of the defining Hamiltonian can be simplified into a sum of harmonic oscillators by decomposing the fields in a plane wave basis into creation and annihilation operators, in the case of the kink Hamiltonian we will see a simplification by decomposing the fields into a basis of normal modes.  The normal modes of the kink are the constant frequency solutions of the classical equations of motion for $H\p_2$
\beq
\V{2}{\g}(x)=\omega^2{\g}(x)+{\g}^{\prime\prime}(x)\hsp \phi(x,t)=e^{-i\omega t}\g(x). \label{sl}
\eeq
There are three kinds of normal mode, classified by their frequencies.  First, all kinks have a zero mode $\g_B$, which generates their translations and has frequency $\omega_B=0$.  Second, some kinks have real shape modes $\g_S$ each of which has a frequency $\omega_S$ in the range $0<\omega_S<m$.  Finally, for each real number $k$ there will be an unbound meson mode $\g_k$ with frequency $\omega_k=\sqrt{m^2+k^2}$.    

As Eq.~(\ref{sl}) is a Sturm-Liouville equation, these modes are a basis of the space of functions of $x$.  We will thus define the normalization of the normal modes, and the separation into $\g_k$ and $\g_{-k}$, by asserting $\g^*_k=\g_{-k}$ and 
\beq
\int dx |{\g}_{B}(x)|^2=1,\
\int dx {\g}_{k_1} (x) {\g}^*_{k_2}(x)=2\pi \delta(k_1-k_2),\ 
\int dx {\g}_{S_1}(x){\g}^*_{S_2}(x)=\delta_{S_1S_2}. \label{comp}
\eeq

As the Schrodinger picture fields $\phi(x)$ and $\pi(x)$ only depend on $x$, they may be decomposed in this basis \cite{cahill76}
\bea
\phi(x)&=&\phi_0 {\g}_B(x) +\ppin{k}\left(B_k^\ddag+\frac{B_{-k}}{2\omega_k}\right) {\g}_k(x)\label{bdec}\hsp
B^\ddag_k=\frac{B^\dag_k}{2\omega_k}\\
\pi(x)&=&\pi_0 {\g}_B(x)+i\ppin{k}\left(\omega_kB_k^\ddag - \frac{B_{-k}}{2}\right) {\g}_k(x)\hsp B_{-S}=B_S\nonumber
\eea
where $\dint dk/(2\pi)$ contains a sum over shape modes $S$ and the coefficients satisfy the algebra
\beq
{[\phi_0,\pi_0]}=i\hsp
[B_{S_1},B^\ddag_{S_2}]=\delta_{S_1S_2}\hsp
[B_{k_1},B^\ddag_{k_2}]=2\pi\delta(k_1-k_2).\nonumber
\eeq

In this basis, the free kink Hamiltonian $H\p_2$ takes the simple form
\beq
H\p_2=Q_1+\hpt\hsp \hpt=\frac{\pi_0^2}{2}+\ppin{k}\ok{} B^\ddag_k B_k \label{hpt}
\eeq
where $Q_1$ is the one-loop correction to the kink mass.  The first term in $\hpt$ describes the quantum mechanics of a free particle, which in this case represents the center of mass motion of the kink.  The second term is a sum of quantum harmonic oscillators, one for each continuum mode $k$ and each shape mode $S$.  Thus, the free theory is a sum of solved quantum mechanical systems.  One can then immediately write the ground state $\vac_0$ of $H\p_2$, it is the simultaneous ground state of each of these subsystems
\beq
\pi_0\vac_0=B_k\vac_0=B_S\vac_0=0. \label{v0}
\eeq
This is the first approximation, in the kink frame, to the kink ground state $\vac$ in our semiclassical expansion.  The corresponding first approximation to the kink ground state in the defining frame is $\df\vac_0$.

What about the interaction terms?  We will define normal-mode normal ordering $::_b$ by placing all $\pi_0,\ B_S$\ and $B_k$ on the right of all $\phi_0,\ B^\ddag_S$\ and $B^\ddag_k$.  In light of (\ref{v0}), normal-mode normal ordering is useful as many terms annihilate the free vacuum $\vac_0$.  Then a Wick's theorem \cite{mewick} relates plane-wave normal ordering $::_a$ to normal-mode normal ordering $::_b$.  We will only be interested in $H\p_3$ and so we will only need Wick's theorem in the cubic case, where it is
\beq
:\phi^3(x):_a=:\phi^3(x):_b + 3\I(x)\phi(x)\hsp
\I(x)=\pin{k}\frac{\left|{\g}_{k}(x)\right|^2-1}{2\omega_k}+\sum_S \frac{\left|{\g}_{S}(x)\right|^2}{2\omega_k}.\label{di}
\eeq
Thus the leading interaction $H\p_3$ can be written
\beq
H\p_{3}=\lambda^{1/2}\int dx\frac{V^{(3)}(\sqrt{\lambda}f(x))}{6}\left(:\phi^3(x):_b+3\I(x)\phi(x)\right). \label{h3}
\eeq
We will not need the $\I(x)$ term in the tree-level calculation presented below, but it is relevant at higher orders.

\section{Twice-Excited Shape Mode's Decay} \label{due}

\subsection{The Twice-Excited Shape Mode}

In the kink frame, the leading order ground state $\vac_0$ is annihilated by the free kink Hamiltonian
\beq
\hpt\vac_0=0. \label{vacz}
\eeq
The operator $B^\ddag_S$ excites the shape mode.  At leading order, the once-excited shape mode
\beq
|1\rangle=B^\ddag_S\vac_0
\eeq
has energy given by the frequency $\os$ of the shape mode
\beq
\hpt|1\rangle=\os|1\rangle.
\eeq
This frequency is necessarily less than the continuum mass threshold
\beq
\os<m
\eeq
as the shape mode is by definition bound to the kink, and so the once-excited shape mode state $|1\rangle$ is stable.

On the other hand, the twice-excited shape mode $|2\rangle$ has twice the energy of $|1\rangle$
\beq
|2\rangle=B^\ddag_SB^\ddag_S\vac_0\hsp
\hpt|2\rangle=2\os|2\rangle.
\eeq
Let us a consider a model such that
\beq
2\os>m.
\eeq
Then a kink with a twice-excited shape mode can decay into a ground state kink plus a single continuum meson excitation.   
\beq
|2\rangle\stackrel{time}{\longrightarrow}|k\rangle\hsp 
|k\rangle=B^\ddag_k\vac_0\hsp
\hpt |k\rangle=\ok{}|k\rangle.
\eeq
This decay is on-shell if
\beq
k=\pm k_I\hsp
\eeq
where we define $k_I$ by
\beq
\omega_{k_I}=2\omega_S\hsp
k_I>0. \label{guscio}
\eeq
However, as is usual in the derivation of Fermi's Golden Rule, it is convenient not to impose that the final state be on-shell as very small deviations from the on-shell condition, of the order allowed by the uncertainty principle, are necessary to obtain the correct scaling of the decay rate with respect to the density of states.

Note that the twice-excited kink $|2\rangle$ cannot decay to the state $B^\ddag_k|1\rangle$ with a once-excited kink and an on-shell meson, because conservation of energy demands that the meson would have energy $\omega_S$, which is less than the continuum threshold $m$.  In this sense the transition $|2\rangle\rightarrow|1\rangle$ is forbidden.

\subsection{Time Evolution}

In the Schrodinger picture of quantum field theory, any state can be evolved from time $0$ to time $t$ by acting with the time evolution operator, which in the vacuum frame is $e^{-iHt}$.  In the kink frame, the time evolution operator is $e^{-iH\p t}$
\beq
e^{-iH\p t}|\rm{time\ 0}\rangle=|\rm{time\ }t\rangle.
\eeq
The operator $H\p$ contains many terms.  For example it contains the scalar $Q$ which is the mass of the kink in its ground state.  This contributes a total phase to the kink state which, for a calculation in the one-kink sector, has no observable consequences.  Therefore, for simplicity, we will drop it.

Once the scalar has been dropped, the leading term in $H\p$ is the free Hamiltonian $\hpt$ given in Eq.~(\ref{hpt}).   The only interaction term that will be relevant for the leading order of the decay of two shape modes is contained in the first term in Eq.~(\ref{h3})
\beq
H_I=\frac{\sqrt{\lambda}}{8\os^2}\pin{k}V_{SSk} B_k^\ddag B_S^2\hsp
V_{SSk}=\int dx \vt \g_k(x) \g_s^2(x). \label{hi}
\eeq
$H_I$ converts two shape modes into a continuum mode
\beq
H_I|2\rangle=\frac{\sqrt{\lambda}}{4\os^2}\pin{k}V_{SSk}|k\rangle.
\eeq

Now we are ready to describe the time evolution of the twice-excited kink state $|2\rangle$.  Under the free Hamiltonian $\hpt$ it evolves via a trivial phase rotation
\beq
e^{-i\hpt t}|2\rangle=e^{-2i\os t}|2\rangle.
\eeq
At order $O(\lambda)$ we may expand the evolution operator, keeping only terms which contain precisely one power of $H_I$
\bea
e^{-i\left(\hpt+H_I\right) t}|_{O(\sqrt{\lambda})}|2\rangle
&=&\sum_{n=1}^{\infty}\frac{(-it)^n}{n!}\left(\hpt+H_I\right)^n|_{O(\sqrt{\lambda})}|2\rangle\\
&=&\sum_{n=1}^{\infty}\frac{(-it)^n}{n!}\sum_{m=0}^{n-1}\hpt{}^m H_I \hpt{}^{n-m-1}|2\rangle\nonumber\\
&=&\frac{\sqrt{\lambda}}{4\os^2}\pin{k}V_{SSk} \sum_{n=1}^{\infty}\frac{(-it)^n}{n!}\sum_{m=0}^{n-1}\ok{}^m (2\os)^{n-m-1}|k\rangle
\nonumber\\
&=&\frac{\sqrt{\lambda}}{8\os^3}\pin{k}V_{SSk} \sum_{n=1}^{\infty}\frac{(-2i\os t)^n}{n!}\frac{1-(\ok{}/(2\os))^n}{1-\ok{}/(2\os)}|k\rangle
\nonumber\\
&=&
-\frac{i\sqrt{\lambda}}{2\os^2}\pin{k}V_{SSk}e^{-i(\os+\ok{}/2) t}\frac{\sin\left((\ok{}/2-\os)t\right)}{\ok{}-2\os}|k\rangle.
\nonumber
\eea

The matrix element with respect to
\beq
\langle k|={}_0\langle 0|\frac{B_k}{2\ok{}}
\eeq
is easily calculated.  Using
\beq
\langle k|k\p\rangle=\frac{{}_0\langle 0\vac_0}{2\ok{}} 2\pi\delta(k\p-k)
\eeq
one finds
\beq
\langle k|e^{-i\left(\hpt+H_I\right) t}|2\rangle|_{O(\sqrt{\lambda})}=-\frac{i\sqrt{\lambda}}{4\os^2\ok{}}V_{SSk}e^{-i(\os+\ok{}/2) t}\frac{\sin\left((\ok{}/2-\os)t\right)}{\ok{}-2\os}{}_0\langle 0|0\rangle_0. \label{matelt}
\eeq

\subsection{The Decay Rate}

Let $\mathcal{P}$ be a projection from the Hilbert space of states to some subspace.  Then the probability $P$ that a state $|\psi\rangle$, when measured, is found to be in the subspace is
\beq
P=\frac{\langle\psi|\mathcal{P}|\psi\rangle}{\langle\psi|\psi\rangle}. \label{prob}
\eeq
We are interested in the probability that the shape modes decay to a continuum mode.   Therefore the subspace should be the subspace of one-meson states in the one-kink sector, which is generated by the states $|k\rangle$.  Therefore $\mathcal{P}$ satisfies
\beq
\mathcal{P}|k\rangle=|k\rangle.
\eeq
This fixes the normalization, so that
\beq
\mathcal{P}=\pin{k}\frac{2\ok{}}{{}_0\langle 0|0\rangle_0} |k\rangle\langle k|.
\eeq

We are interested in the state found at time $t$ in the previous subsection
\beq
|\psi\rangle=e^{-i(\hpt+H_I) t}|2\rangle|_{O(\sqrt{\lambda})}.
\eeq
Since the evolution operator is unitary, it does not change the norm and so
\beq
\langle\psi|\psi\rangle=\langle 2|2\rangle=\frac{{}_0\langle 0\vac_0}{2\os^2} .
\eeq
Putting this together with the matrix elements (\ref{matelt}), the decay probability at time $t$ is
\bea
P(t)&=&\pin{k} 4\os^2\ok{} \frac{\langle\psi|k\rangle\langle k |\psi\rangle}{{}_0\langle 0\vac_0^2}\\
&=&\frac{\lambda}{4\os^2}\pin{k} \frac{\left|V_{SSk}\right|^2}{\ok{}}\frac{\sin^2\left((\ok{}/2-\os)t\right)}{\left(\ok{}-2\os\right)^2}
\nonumber
\eea
corresponding to a decay rate of
\beq
\dot{P}(t)=\frac{\lambda}{8\os^2}\pin{k} \frac{\left|V_{SSk}\right|^2}{\ok{}}\frac{\sin\left((\ok{}-2\os)t\right)}{\left(\ok{}-2\os\right)} \label{pd}.
\eeq

\subsection{The Mean Lifetime}

The result (\ref{pd}) for the tree-level decay rate is exact at all times $t$.  At very small times, as $t\rightarrow 0$, the decay rate tends linearly to zero.  This is a manifestation of the quantum Zeno effect.  We will instead be interested in times $t$ in the window
\beq
\frac{1}{m} << t << O(\lambda^{-1}). \label{tlim}
\eeq
The lower limit ensures that that the decay has passed the quantum Zeno phase, and entered the exponential decay phase.  The upper limit is necessary for the tree-level contribution discussed here to dominate the decay rate.  For example, at much larger times there would be a reduction in the decay rate reflecting the fact that an appreciable part of the originally excited shape mode has already decayed.  The upper bound on $t$ in Eq.~(\ref{tlim}) is proportional to $1/\lambda$ and also to some model-dependent function of the mass scales in the model.

The lower bound on the time implies that the $sinc$ function is in a regime where it is a nascent $\delta$ function
\beq
\stackrel{\rm{lim}}{{}_{t\rightarrow\infty}}\pin{k} F(k) \frac{\sin(\alpha (k-k_I) t)}{\alpha (k-k_I) }=\frac{F(k_I)}{2\alpha}\rm{sign}(\alpha) \label{nasc}
\eeq
for any nonzero $\alpha$.  In particular, as at large times $k$ will be close to one of the two on-shell values $\pm k_I$, we may expand
\beq
\ok{}-2\os=\frac{\partial\ok{}}{\partial k}|_{k=\pm k_I}(k \mp k_I)=\frac{\pm k_I}{\ok{I}}(k \mp k_I)
\eeq
up to corrections of order $O((k \mp k_I)^2)$.  Choosing one sign $\pm k_I$, these corrections are large near $\mp k_I$.  Therefore to reduce the integral in (\ref{pd}) to the form (\ref{nasc}) it is necessary to sum over the two on-shell values $\pm k_I$.

For the value $\pm k_I$ we identify
\beq
F(k)=\frac{\left|V_{SSk}\right|^2}{\ok{}}\hsp \alpha=\frac{\pm k_I}{\ok{I}}
\eeq
and so, for $t>>1/m$, the decay rate tends to a constant
\beq
\dot{P}(t)=\frac{\lambda}{16\os^2} \frac{\left|V_{SSk_I}\right|^2+\left|V_{SS-k_I}\right|^2}{k_I}.
\eeq
In particular, for a symmetric potential, with an antisymmetric kink, this is simply
\beq
\dot{P}(t)=\frac{\lambda}{8\os^2} \frac{\left|V_{SSk_I}\right|^2}{k_I}. \label{pgen}
\eeq
The power radiated $\dot{E}$ and the mean lifetime $\tau$ are then
\beq
\dot{E}=2\os \dot{P}=\frac{\lambda}{4\os} \frac{\left|V_{SSk_I}\right|^2}{k_I}\hsp
\tau=\frac{1}{\dot{P}}=\frac{8\os^2}{\lambda} \frac{k_I}{\left|V_{SSk_I}\right|^2}. \label{egen}
\eeq

One may compare our main result (\ref{pgen}) with expectations from Fermi's Golden Rule.  The normalized matrix element for the transition to $k=\pm k_I$ is
\beq
A_\pm=\frac{\langle \pm k_I|H_I|2\rangle}{\sqrt{\langle\pm k_I|\pm k_I\rangle\langle 2|2\rangle}}=\frac{\sqrt{\lambda}V_{SS\pm k_I}/(8\omega_S^2\ok{I})}{\sqrt{2\pi}/(2\omega_S\sqrt{\ok{I}})}=.\frac{\sqrt{\lambda}V_{SS\pm k_I}}{4\omega_S\sqrt{2\pi\ok{I}}}.
\eeq
Using the fact that the density of states is $\ok{I}/k_I$ and summing over the two signs of the final momentum, Fermi's Golden Rule yields
\beq
2\pi \frac{\ok{I}}{k_I}\sum_{\pm}|A_\pm|^2=\frac{\lambda\left|V_{SS\pm k_I}\right|^2}{8\omega_S^2 k_I}
\eeq
which indeed agrees with the transition rate reported in Eq.~(\ref{pgen}).

\subsection{Example: The $\phi^4$ Kink}

The $\phi^4$ double-well model is described by the potential
\beq
V(\sqrt{\lambda}\phi)=\frac{\lambda \phi^2}{4}\left(\sqrt{\lambda}\phi(x)-\beta\sqrt{8}\right)^2
\eeq
where $\beta$ is a mass scale, related to the scalar mass $m$ and the shape mode frequency $\os$ by
\beq
m=2\beta\hsp \os=\sqrt{3}\beta.
\eeq
The on-shell condition (\ref{guscio}) then leads to a wave number and frequency of
\beq
k_I=2\sqrt{2} \beta\hsp
\ok{I}=2\sqrt{3}\beta   \label{k4}
\eeq
for the emitted radiation \cite{mm}.

In Ref.~\cite{phi42loop} it was found that
\beq
V_{SSk}=i\pi\frac{3}{8\sqrt{2}}\frac{k^2\ok{}(2\beta^2-k^2)}{\beta^3\sqrt{\beta^2+k^2}}\ck \label{v4}
\eeq
and so, on-shell,
\beq
V_{SSk_I}=-6\sqrt{6}i\pi\beta\csch(\pi\sqrt{2}).
\eeq
Substituting this into our general formulas we find
\beq
\dot{P}=\frac{9\pi^2\csch^2(\pi\sqrt{2})}{2\sqrt{2}} \frac{\lambda}{\beta}\hsp
\dot{E}=\frac{9\sqrt{3}\pi^2\csch^2(\pi\sqrt{2})}{\sqrt{2}} \lambda \hsp
\tau=\frac{2\sqrt{2}}{9\pi^2\csch^2(\pi\sqrt{2})} \frac{\beta}{\lambda}.
\eeq

\section{Coherent Excited Shape Mode's Decay} \label{tanti}

In the Appendix of Ref.~\cite{mm}, the authors calculated the decay rate of a shape-mode excitation on a $\phi^4$ kink in classical field theory.  As a consistency check on our formalism, we will calculate the decay rate of the same state in quantum field theory.  Of course the quantum calculation is more complicated and yields the same answer.  However, in principle it may be extended to the next order where it will reveal corrections inaccessible to classical field theory.

\subsection{Coherent States}

The first ingredient that we will need is the quantum state corresponding to the classical kink with an excited shape mode.   This classical solution is not stationary, even ignoring the fact that the mode decays into radiation, and so the corresponding state will not be an eigenstate of the free Hamiltonian $\hpt$.  The state at one time during this oscillation was given in Ref.~\cite{chris}, in the case of a wave packet.  Here we consider states which are invariant under spatial translations, not localized wave packets, but our coherent state will nonetheless be similar
\beq
|\alpha,t\rangle=\rm{exp}\left[\alpha\left(\sqrt{2\os}B^\ddag_S e^{-i\os t}-\frac{B_S}{\sqrt{2\os}}e^{i\os t}\right)\right]\vac_0. \label{cs}
\eeq
Note that the exponentiated quantity is antihermitian and so the exponential operator is unitary, implying that it does not affect the normalization of the state
\beq
\langle \alpha,t|\alpha,t\rangle={}_0\langle 0\vac_0.
\eeq

Let us check to see if these correspond to the correct classical states.  Using the commutators
\bea
\left[B_S^\ddag,e^{\left[\alpha\left(\sqrt{2\os}B^\ddag_S e^{-i\os t}-\frac{B_S}{\sqrt{2\os}}e^{i\os t}\right)\right]}\right]&=&\frac{\alpha}{\sqrt{2\os}}e^{i\os t}e^{\left[\alpha\left(\sqrt{2\os}B^\ddag_S e^{-i\os t}-\frac{B_S}{\sqrt{2\os}}e^{i\os t}\right)\right]}\\
\left[B_S,e^{\left[\alpha\left(\sqrt{2\os}B^\ddag_S e^{-i\os t}-\frac{B_S}{\sqrt{2\os}}e^{i\os t}\right)\right]}\right]&=&{\alpha}{\sqrt{2\os}}e^{-i\os t}e^{\left[\alpha\left(\sqrt{2\os}B^\ddag_S e^{-i\os t}-\frac{B_S}{\sqrt{2\os}}e^{i\os t}\right)\right]}\nonumber
\eea
together with the decompositions (\ref{bdec}), which include the shape mode terms
\beq
\phi(x)\supset \left(B^\ddag_S+\frac{B_S}{2\os}\right)\g_S(x)\hsp
\pi(x)\supset i\left(\os B^\ddag_S-\frac{B_S}{2}\right)\g_S(x)
\eeq
one arrives at the expectation values
\beq
\frac{\langle \alpha,t|\phi(x)|\alpha,t\rangle}{\langle \alpha,t|\alpha,t\rangle}=\alpha\sqrt{\frac{2}{\os}}\g_S(x) \cos\left(\os t\right)\hsp
\frac{\langle \alpha,t|\pi(x)|\alpha,t\rangle}{\langle \alpha,t|\alpha,t\rangle}=-\alpha\sqrt{2\os}\g_S(x) \sin\left(\os t\right). \label{vevs}
\eeq
Recall that we are working in the kink frame, and so the first expectation value is the deviation of the classical field from the classical kink solution $\phi(x,t)=f(x)$.  One thus finds that this state represents a classical field configuration with an excited shape mode whose amplitude is $\alpha\sqrt{2/\os}$, as was found using a matching to the action on moduli space in Ref.~\cite{chris}.

Before studying the decay of the shape mode, which results from the interaction $H_I$, we should understand the oscillations of the shape mode, which are described by the free Hamiltonian $\hpt$.  Note that any functional of $\hpt$, when it is moved to the right past a $B^\ddag_S$ or $B_S$, becomes the same functional with $\hpt$ replaced by $\hpt+\os$ or $\hpt-\os$ respectively.  In particular, it is easy to show that
\beq
e^{-i\hpt t}B^\ddag_S=B^\ddag_S e^{-i(\hpt+\os) t}\hsp
e^{-i\hpt t}B_S=B_S e^{-i(\hpt-\os) t}.
\eeq
This allows us to move the free evolution operator $e^{-i\hpt t}$ past any function $F$ of $B^\ddag_S$ and $B_S$
\beq
e^{-i\hpt t}F[B^\ddag_S,B_S]=F[B^\ddag_Se^{-i\os t},B_Se^{i\os t}]e^{-i\hpt t}.
\eeq

Setting $F$ to be the exponential in the coherent state (\ref{cs}), and using (\ref{vacz}) one finds
\beq
e^{-i\hpt t}|\alpha,0\rangle=|\alpha,t\rangle. \label{freeev}
\eeq
Therefore the classical oscillation of the shape mode is described by the evolution under the free part $\hpt$ of the kink Hamiltonian $H\p$.  Any decay of the shape mode can only be described by the interaction terms in $H\p$.

\subsection{Time Evolution}

The coherent states $|\alpha,t\rangle$ are eigenstates of the shape mode annihilation operator
\beq
B_S|\alpha,t\rangle=\sqrt{2\os}e^{-i \os t}\alpha|\alpha,t\rangle.
\eeq
Let us define a state $|\alpha,t,k\rangle$ consisting of a coherent shape mode excitation plus a single free meson
\beq
|\alpha,t,k\rangle=B^\ddag_k|\alpha,t\rangle\hsp \langle\alpha,t,k|\alpha,t,k\p\rangle=\frac{2\pi\delta(k-k\p)}{2\ok{}}{}_0\langle 0\vac_0.
\eeq
Putting this all together, the action of the interaction $H_I$, defined in Eq.~(\ref{hi}), on a coherent state is
\beq
H_I|\alpha,t\rangle=\frac{\sqrt{\lambda}\alpha^2e^{-2i \os t}}{4\os}\pin{k}V_{SSk}|\alpha,t,k\rangle.
\eeq
The free evolution of the shape mode plus meson state is
\beq
e^{-i\hpt t_2}|\alpha,t_1,k\rangle=e^{-i\ok{}t_2}|\alpha,t_1+t_2,k\rangle.
\eeq

To evolve the state $|\alpha,0\rangle$ we will use the identity
\beq
e^{-i(\hpt+H_I)t}=e^{-i\hpt t}-i\int_0^t dt_1 e^{-i\hpt (t-t_1)} H_1 e^{-i\hpt t_1}+O(H_1^2).
\eeq
This is easily checked by differentiating both sides with respect to $t$, and the constant of integration can be checked at $t=0$.    Keeping\footnote{We drop the dominant term, of order $O(\lambda^0)$, which was already reported in Eq.~(\ref{freeev}) as it will not contribute to the matrix elements below.} only terms of $O(\sqrt{\lambda})$
\bea
e^{-i(\hpt+H_I)t}|\alpha,0\rangle
&=&-i\int_0^t dt_1 \frac{\sqrt{\lambda}\alpha^2e^{-2i \os t_1-i\ok{}(t-t_1)}}{4\os}\pin{k}V_{SSk}|\alpha,t,k\rangle\\
&=&\frac{-i\sqrt{\lambda}\alpha^2}{4\os}\pin{k} V_{SSk}e^{-i(\os+\ok{}/2)t}\frac{\rm{sin}\left(\left(\os-\ok{}/2\right)t\right)}{\os-\ok{}/2}|\alpha,t,k\rangle.\nonumber
\eea
This is described by the matrix elements
\beq
\langle\alpha,t,k|e^{-i(\hpt+H_I)t}|\alpha,0\rangle=\frac{-i\sqrt{\lambda}\alpha^2}{8\os\ok{}} V_{SSk}e^{-i(\os+\ok{}/2)t}\frac{\rm{sin}\left(\left(\os-\ok{}/2\right)t\right)}{\os-\ok{}/2}{}_0\langle 0\vac_0.
\eeq
It is tempting to compare these matrix elements directly with Eq.~(\ref{matelt}), but one must recall that the two initial states $|2\rangle$ and $|\alpha,0\rangle$ have different normalizations.

\subsection{The Decay Rate}

Again we would like to compute the probability that the state, at time $t$, is in some subspace of the Hilbert space.  A single decay process will place the state in the subspace generated by the states $|\alpha,t,k\rangle$.  The projector
\beq
\mathcal{P}=\pin{k}\frac{2\ok{}}{{}_0\langle 0|0\rangle_0} |\alpha,t,k\rangle\langle \alpha,t,k| \label{projb}
\eeq
preserves this subspace.  

Substituting (\ref{projb}) into the general formula (\ref{prob}) for the probability that the state
\beq
|\psi\rangle=e^{-i(\hpt+H_I)t}|\alpha,0\rangle\hsp
\langle \psi|\psi\rangle={}_0\langle 0\vac_0
\eeq
is in the subspace, we find
\bea
P(t)&=&\pin{k} 2\ok{} \frac{\langle\psi|\alpha,t,k\rangle\langle \alpha,t,k |\psi\rangle}{{}_0\langle 0\vac_0^2}\\
&=&\frac{\lambda\alpha^4}{8\os^2}\pin{k} \frac{\left|V_{SSk}\right|^2}{\ok{}}\frac{\sin^2\left((\ok{}/2-\os)t\right)}{\left(\ok{}-2\os\right)^2}
\nonumber
\eea
corresponding to a decay rate of
\beq
\dot{P}(t)=\frac{\lambda\alpha^4}{16\os^2}\pin{k} \frac{\left|V_{SSk}\right|^2}{\ok{}}\frac{\sin\left((\ok{}-2\os)t\right)}{\left(\ok{}-2\os\right)} \label{pdc}.
\eeq

This is an observable quantity, independent of normalization parameters, and so it may be directly compared with the twice-excited shape mode of Eq.~(\ref{pd}).  One finds that the decay rate in the coherent case is greater than the twice-excited case by a factor of $\alpha^4/2$.  This is not so surprising, as the decay requires a choice of 2 shape modes.  In the twice-excited kink there was only one such choice.  The coherent state, on the other hand, is not an eigenstate of the shape mode number operator.  However, the expected number of shape modes scales is of order $O(\alpha)$, and so the number of pairs is of order $O(\alpha^2)$.  One thus expects the amplitude and the rate to be of orders $O(\alpha^2)$ and $O(\alpha^4)$ respectively.

As the rates (\ref{pd}) and (\ref{pdc}) are related by a simple constant of proportionality, the $k$ integration may be performed as in the twice-excited case, simply keeping this constant.    One therefore finds that at times much greater than the Zeno time, the decay rate is
\beq
\dot{P}(t)=\frac{\lambda\alpha^4}{16\os^2} \frac{\left|V_{SSk_I}\right|^2}{k_I}
\eeq
and the power radiated $\dot{E}$ is
\beq
\dot{E}=2\os \dot{P}=\frac{\lambda\alpha^4}{8\os} \frac{\left|V_{SSk_I}\right|^2}{k_I}. \label{rad}
\eeq

The lifetime is no longer given by the inverse rate, as a single decay is no longer sufficient to completely relax the shape mode.  Instead the lifetime, defined to be the exponential decay constant of the energy in the shape mode, can be computed as follows.   First, substituting the classical configuration (\ref{vevs}) into the kink Hamiltonian, one finds the expected energy stored in the shape mode
\beq
E=2\os \alpha^2.
\eeq
Then the lifetime is
\beq
\tau=\frac{E}{\dot{E}}=\frac{16\os^2} {\lambda\alpha^2}\frac{k_I}{\left|V_{SSk_I}\right|^2}.
\eeq
It is equal to the expected lifetime (\ref{egen}) of the twice-excited kink when $\alpha=2$.  On the other hand, a strongly excited shape mode $\alpha>2$ decays more gradually, although the absolute radiation output (\ref{rad}) is higher.

\subsection{Example: The $\phi^4$ Model}

In the case of the $\phi^4$ double-well model, substituting Eqs.~(\ref{k4}) and (\ref{v4}) into (\ref{rad}) one finds that the radiated power is
\beq
\dot{E}=\frac{9\sqrt{3}\pi^2}{2\sqrt{2}}\csch^2(\pi\sqrt{2}) \alpha^4 \lambda . 
\eeq
This agrees with Eq.~(A.20) of Ref.~\cite{mm}, where the authors use units such that
\beq
\lambda=2\hsp
\beta=1\hsp
\alpha=\frac{A_0}{{3^{1/4}}}.
\eeq

\section{Remarks}

We have calculated the tree-level decay rate for the twice-excited kink state and a coherent state corresponding to a classical solution.  This is tree level with respect to the kink Hamiltonian $H\p$, but would have been a nonperturbative calculation had we relied on the defining Hamiltonian $H$.

Of course, the strength of this formalism is that one may go to higher orders with less complexity than traditional approaches, such as the collective coordinate method of Refs.~\cite{gjscc,gj76}.  So, can one go to higher orders?

At higher orders there will be ambiguities to contend with, related to the choice of initial state \cite{decad0,decad1}.  The choice of initial state depends on how the excited kink arose.  An excited kink could result from kink-meson scattering, or from a cosmological phase transition \cite{mmnum}.

In the case of the coherent state the ambiguity is clear, as many different quantum states correspond to the same classical state, yet the choice will determine the quantum corrections to the lifetime.  Similarly, the twice-excited kink state is not a Hamiltonian eigenstate, and so there is no clear prescription for which quantum corrections to include in the initial configuration.  At tree level we simply used the eigenstate of the free Hamiltonian, but at higher orders there are choices to be made.  Ref.~\cite{decad2} suggests an adiabatic approximation, where one always uses an initial state which is an eigenstate of the free Hamiltonian, and suggests that such a state will yield the correct lifetime at subleading order, but not beyond.

Another complication at higher orders is the presence of zero modes $\phi_0$ in the terms in $H\p$ which appear.  In the calculation at this order, there were factors of ${}_0\langle 0\vac_0$ which canceled in the numerator and denominator.  Formally these factors are infinite, as the kink state is translation-invariant and so not normalizable.  However this naive divergence is easily treated by considering a wave packet in $\phi_0$ and, at the end, taking the limit that the width of the wave packet goes to infinity.  The perturbation theory of the other modes does not converge at large $\phi_0$, but in the above tree-level calculation this does not affect the calculation because $\phi_0$ does not mix with the other terms.  On the other hand, at higher orders, moments of $\phi_0$ will appear and those terms will only have a reasonably convergent semiclassical expansion, in the sense of an asymptotic series, in the case of a narrow wave packet \cite{mepacket}.  Treatment of the decay of a twice-excited translation-invariant kink may also be possible, but one will need to project onto the translation-invariant subspace of the Hilbert space during the calculation.  But even once this is done, the dependence on the arbitrary choice of initial condition remains.

The standard solution to this problem is, at higher orders, to calculate not the ill-defined lifetime, but rather to calculate the width of the corresponding resonance in meson-kink scattering.  We expect elastic meson-kink scattering to have an infinite tower of narrow resonances, one for each integer $n\geq 2$, representing the $n$-excited shape mode state.  The initial condition for this process is well-defined, as the kink and meson can be well-separated and so they can each be taken to be separately Hamiltonian eigenstates.  Therefore the width is well-defined, and its inverse width should agree with the the lifetime for as many orders as the lifetime is also well-defined.  We intend to turn to this scattering problem in the near future.

The treatment of the decay of the coherent shape mode paves the way to yet more physically interesting calculations.  For example, one could use the same strategy to construct a coherent state corresponding to any oscillating solution, such as an oscillon, breather or bion.  One can then directly test the claim in Ref.~\cite{quantosc} that in the quantum theory there are perturbative decay channels which dominate the decay rate with respect to the nonperturbative channel which survives in the classical limit. In higher-dimensional models, the results of such a study would be relevant to various open questions in cosmology \cite{cos}. Needless to say, additional decay channels during quantum kink-antikink collisions could have brutal phenomenological consequences for the resonance windows found in classical field theory~\cite{csw,doreyinterp,col22}.

\section* {Acknowledgement}

\noindent
JE is supported by the CAS Key Research Program of Frontier Sciences grant QYZDY-SSW-SLH006 and the NSFC MianShang grants 11875296 and 11675223.   JE also thanks the Recruitment Program of High-end Foreign Experts for support.  AGMC acknowledges financial support from Xunta de Galicia (Centro singular de investigación de Galicia accreditation 2019-2022), by European Union ERDF, and by the “María de Maeztu” Units of Excellence program MDM-2016-0692 and the Spanish Research State Agency.
AGMC is also grateful to the Spanish Ministry of Science, Innovation and Universities, and the European Social Fund for the funding of his predoctoral research activity (\emph{Ayuda para contratos predoctorales para la formaci\'on de doctores} 2019).

\end{document}